\documentclass[aps,reprint,amsmath, amssymb,groupedaddress,floatfix]{revtex4-1}
\usepackage{natbib}
\usepackage{graphicx}
\usepackage{bm}
\usepackage{color,soul}
\usepackage{hyperref}
\usepackage{float}
\usepackage{csquotes}
\usepackage{lipsum, babel}

\usepackage{color,soul}
\usepackage{hyperref}
\hypersetup{colorlinks=true, urlcolor=blue, citecolor=blue}
\usepackage{longtable}

\begin{document}

\date{\today}

\title{Persistent spin textures in halide perovskites induced by uniaxial stress}
\author{Ravi Kashikar$^{1}$ }
\email{ ravik@usf.edu}
\author{Abduljelili Popoola$^{1}$ }
\author{Sergey Lisenkov$^{1}$}
\author{A. Stroppa$^{2}$}
\author{I. Ponomareva$^{1}$}
\email{iponomar@usf.edu}
\affiliation{1. Department of Physics, University of South Florida, Tampa, Florida 33620, USA}
\affiliation{2. Consiglio Nazionale delle Ricerche, Institute for Superconducting and Innovative Materials and Devices (CNR-SPIN), c/o Department of Physical and Chemical Sciences, University of L'Aquila, Via Vetoio I-67100 Coppito, L'Aquila, Italy}


\begin{abstract}
Persistent spin textures are highly desirable for applications in spintronics as they may allow for long carrier spin lifetimes. However, they are also rare as only four point groups can host such textures, and even for these four groups, the emergence of persistent spin textures requires a delicate balance between coupling parameters, which control the strength of spin-momentum interactions.  We use first-principles density functional simulations to predict the possibility of achieving these desirable spin textures through the application of uniaxial stress. Hybrid organic-inorganic perovskite MPSnBr$_3$ (MP = CH$_3$PH$_3$) is a ferroelectric semiconductor which exhibits persistent spin textures in the near to its conduction band minimum and mostly Rashba  type  in the vicinity of  its valence band maximum. Application of uniaxial stress leads to the gradual evolution of the valence bands spin textures from mostly Rashba type to persistent ones under tensile load and to pure Rashba or persistent ones under compressive load. We also report that the material exhibits flexibility, rubber-like response, and both positive and negative piezoelectric constants. Combination of such properties may create opportunities for a flexible/rubbery spintronic devices. 
\end{abstract}
\thispagestyle{empty}

\flushbottom
\maketitle
Persistent spin textures (PST) are unique spin patterns in the momentum space, for which  spin directions are independent of wave vectors \cite{PST_1, PST_rvw1,PST_rvw2,Tao2018,Tao_2021}. They could provide long carrier spin lifetimes through the persistent spin helix mechanism, which is highly desirable for novel applications, such  as spin field effect transistors\cite{PST_rvw1}.  PST could arise from the balance between Rashba and Dresselhaus interactions and have previously been 
realized in group III-V semiconductor heterostructures; GaAs/AlGaAs  and InGaAs/InAlAs \cite{Koralek2009, PST_rvw1,PST_rvw2}. Furthermore, the possibility of tuning the topology of spin textures, including PST, by means of electric field is another highly desirable property, which has been demonstrated or predicted in some ferroelectric semiconductors\cite{GeTe, GeTe_expt}. Ferroelectrics possess spontaneous polarization, whose direction is reversible by the application of external electric field. Such polarization reversal can be associated with  the space-inversion operation, under which the wave vector $\mathbf k$ changes to $-\mathbf k$ while spin remains invariant. Time-reversal symmetry operation brings $-\mathbf k$ back to  $\mathbf k$
 but flips the spin, thus resulting in the reversal of spin direction through the  polarization reversal\cite{Tao_2021}. Interestingly, the space groups allowing for occurrence of spin textures, including the PSTs,  are often polar ones, so they can co-host ferroelectricity, leading to ferroelectricity Rashba effect co-functionality\cite{Zunger_ferro}. Some examples among inorganic and hybrid organic-inorganic materials include GeTe, Sr$_3$Hf$_2$O$_7$, BA$_2$PbCl$_4$ (BA = benzylammonium), (AMP)PbI$_4$ (AMP = 4-
(aminomethyl)piperidinium), TMCM-MnCl$_3$, TMCM-CdCl$_3$ (TMCM = Trimethylchloromethyl),  MPSnBr$_3$ \cite{BAPbCl_Rashba, GeTe, Stroppa_npj, LU20201211, AMP,TMCM_PRB, Ravi_PRM}.

Acosta \textit{et al.,} classified occurrence of various spin textures in different point groups  through crystal and wave vector point group perspective \cite{Acosta_BZ}. Their analysis predicted that, among polar crystal point groups, Rashba and Dresselhaus spin texture may appear in C$_s$ and C$_{2v}$ wave vector point groups. Therefore, it is reasonable to look into the materials with these two point groups for potential PST candidates \cite{SYmm_PST,Absor_2022}. As an example, the automatically curated first-principles database of ferroelectrics \cite{Database} returns 37 inorganic materials belonging to C$_s$ point group and 92 belonging to C$_{2v}$ point group  out of its 255 inorganic  compounds. However, only few of these materials are experimentally confirmed ferroelectrics. Some examples include (Ca,Sr)$_3$Ti$_2$O$_7$, Sr$_3$Sn$_2$O$_7$  and  Ca$_3$Mn$_2$O$_7$,  \cite{SrTiO, SrSno, CMO,RP_series}.

Another class of materials to look for PST co-existing with ferroelectricity is  hybrid organic-inorganic perovskites, many of which   crystallize in C$_s$ and C$_{2v}$ space groups. For instance, we found 33 compounds with confirmed ferroelectricity which crystallize in these space groups and list them in Supplementary Material.  Some  examples  are given in Table~\ref{T1}. Interestingly, TMCM-CdCl$_3$ and MPSnBr$_3$ are  already predicted to exhibit PST either in conduction band (CB) or valence band (VB) \cite{TMCM_PRB, Ravi_PRM}. Hybrid organic-inorganic perovskites are known for hosting a multitude of properties, such as  optical, ferroelectric and magnetic properties and thus used in photodetectors, photovoltaics, actuators, among others \cite{Stroppabook, HOIP_Rvw_1,HOIP_Rvw_1, Rvw1, Rvw2, PPDs1, siwach2022design}.   They  are relatively soft,  flexible, and may exhibit high anisotropy of their elastic properties\cite{Partha_AMI}, which in turn could open a way to tunability of their electronic properties through mechanical deformations.

\begin{table*}
\centering
\small
\caption{Some representative  ferroelectric hybrid organic-inorganic halide perovskites, their crystal point group, spin-splitting energy, and spin texture around valence band (VB) maximum and conduction band (CB) minimum. The path in the Brillouin zone along which spin splitting is computed is also provided.  }
\begin{tabular}{cccccc}
\hline
Compound & Point & Region&\hspace{1cm} k-path&Spin& Spin texture \\
&  Group & &&splitting (meV)&  \\
\hline
(DMMA)CdCl$_3$ &C$_{2v}$ & VB&\hspace{1cm}S-$\Gamma$-S& 0.1& D-P  \\
 && CB&\hspace{1cm}S-$\Gamma$-S& $\sim$0& R-P   \\
(BE)PbBr$_3$ &C$_{2v}$ & VB&\hspace{1cm}$\Gamma$-S-$\Gamma$ &$\sim$0& P  \\
 && CB&\hspace{1cm}S-$\Gamma$-S& 0.9& R   \\
 (MNCC)CdCl$_3$ &C$_{2v}$ & VB&\hspace{1cm}S-$\Gamma$-S &$\sim$0.1& D-P   \\
 && CB&\hspace{1cm}S-$\Gamma$-S& $\sim$0& R-P   \\
 
  (TMCM)CdCl$_3$ &C$_{s}$ & VB&\hspace{1cm}Y-$\Gamma$-Y& 0.2& P  \\
 && CB&\hspace{1cm}Y-$\Gamma$-Y& $\sim$0& R-P  \\
  MPSnBr$_3$ &C$_{2v}$ & VB&\hspace{1cm}Y-$\Gamma$-Y& $\sim$0& R  \\
 && CB&\hspace{1cm}S-$\Gamma$-S& 0.3& R-P   \\
  MPSnBr$_3$ &C$_{s}$ & VB&\hspace{1cm}$\Gamma$-X-$\Gamma$ &1.9& R-P   \\
 && CB&\hspace{1cm}$\Gamma$-X-$\Gamma$& 13.6& P   \\
\hline
\multicolumn{5}{c}{\footnotesize DMMA=N,N-dimethylallylammonium, BE=BrCH$_2$CH$_2$N-(CH$_3$)$_3$],}\\
\multicolumn{5}{c}{\footnotesize 
MNCC=[Me$_3$NCH$_2$ CH$_2$OH], TMCM=Trimethylchloromethyl, MP = CH$_3$PH$_3$. }\\
\multicolumn{5}{c}{\footnotesize 
The structural data of these perovskites were obtained from Ref. \cite{DMMA, BEPbBr,MNCC,TMCM_PRB,MPSnBr}}
\end{tabular}
\label{T1}
\end{table*}

\begin{figure}
\centering
\includegraphics[width=0.5\textwidth]{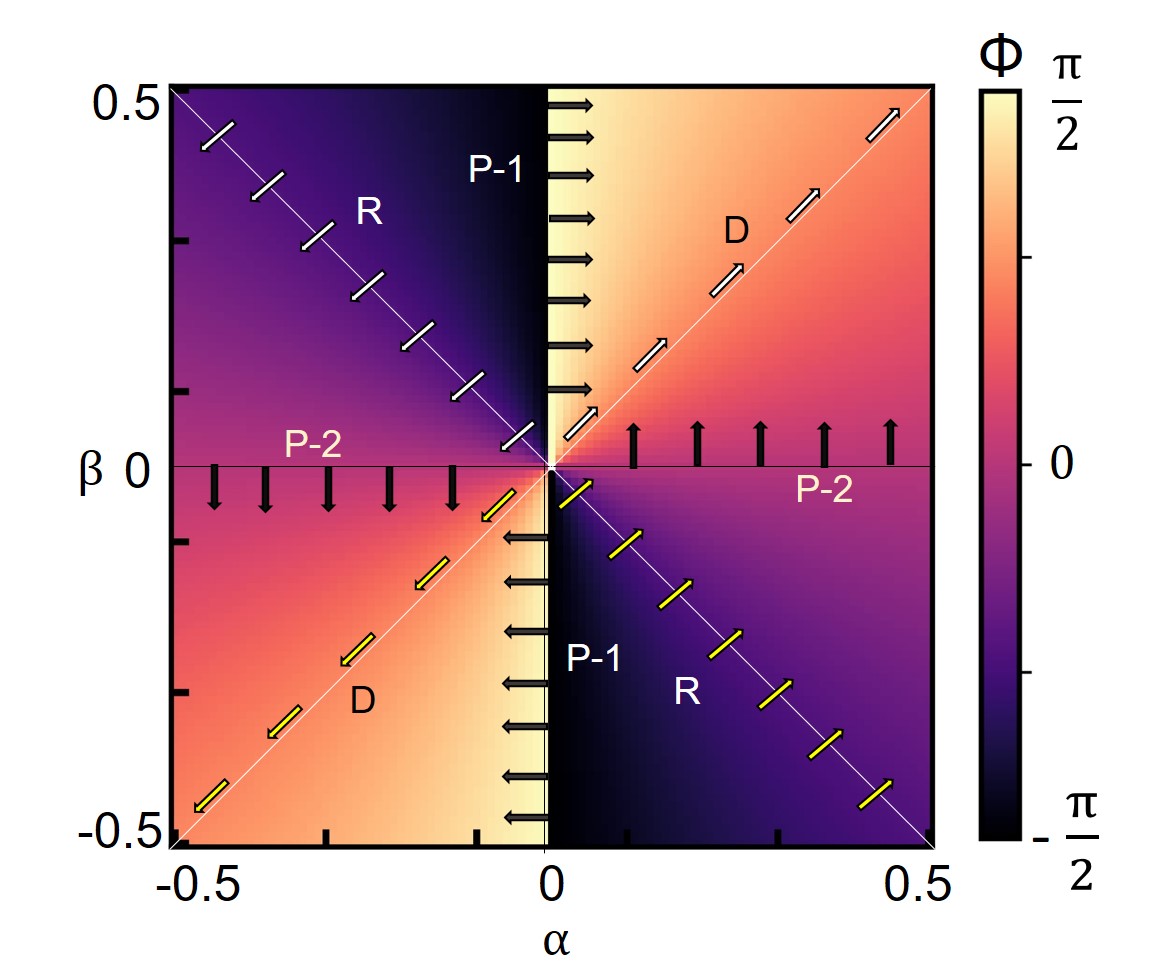}
\caption{ Spin texture types diagram for the effective Hamiltonian (Eq.~\ref{e1}) when $\delta=\gamma=$0. The arrows give orientation of the spins. R and D label represents Rashba and Dresselhaus spin texture types, while  P-1 and P-2 labels are for PST where spin directions are along k$_x$ and k$_y$ directions, respectively.    }
\label{fig1}
\end{figure}

 Table~\ref{T1} contains a list of some experimentally synthesized ferroelectric hybrid organic-inorganic perovskites that are members of C$_s$ and C$_{2v}$ crystal point groups. To estimate their spin-splitting we used first principle based density functional theory (DFT) based calculations as implemented in Vienna Ab-initio Simulation Package (VASP) \cite{VASP}. Perdew-Burke-Ernzerhof (PBE) exchange-correlation functional and projector-augmented-wave (PAW) pseudopotentials with dispersion correction as proposed by Grimme \textit{et al}.\cite{Bloch,Grimme} have been utilized as they have previously been shown to  predict properties that are in good agreement with experimental data \cite{D1,D2}. The energy cutoff was set  to  700 eV, which is 1.3 of the largest cutoff for the elements listed in the pseudopotential file. For  k-point mesh, we chose uniform kpoint density, in the range  0.18-0.23 \AA$^{-1}$. The energy convergence for electronic structure calculation was set to 0.001 meV. The electronic structure calculations incorporate spin-orbit coupling (SOC) and are carried out  on experimental crystal structures. The computed band structures and spin textures are reported in Supplementary Material (Figs.~\ref{figS1} and \ref{figS2}). 
 
\begin{figure*}
\centering
\includegraphics[width=0.7\textwidth]{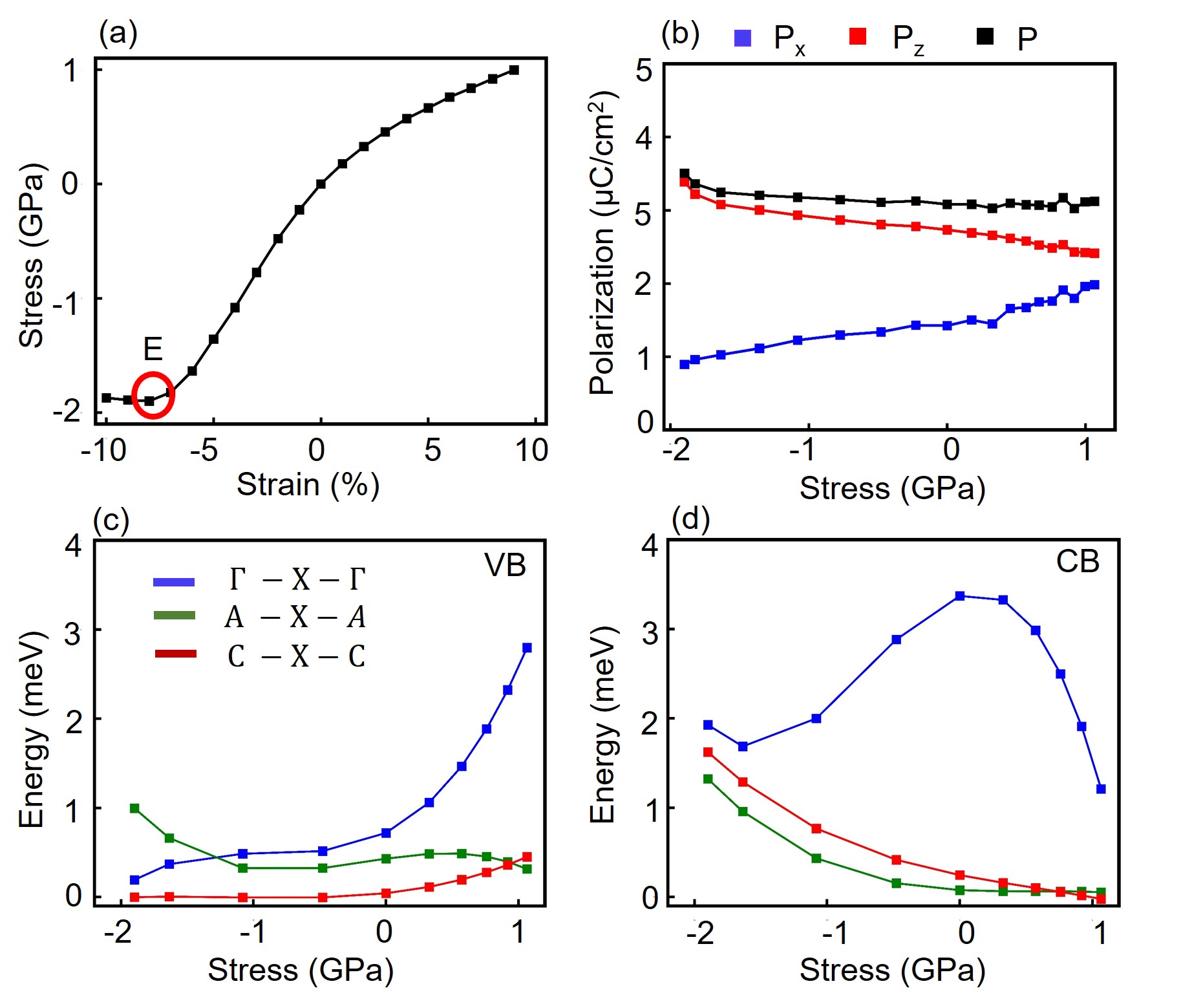}
\caption{ (a) Stress-strain dependence  for MPSnBr$_3$. The red circle indicates elastic limit.  (b) Polarization components and magnitude as a function of uniaxial stress. (c) and (d) Spin splitting strengths as a function of uniaxial stress for VB and CB.}
\label{fig2}
\end{figure*}

 Table~\ref{T1} summarizes our computational data for spin splitting in different bands and type of spin textures in the vicinity of valence band maximum (VBM) and conduction band minimum (CBM).  The calculations predict that most of them indeed exhibit spin splitting, which, however, is rather small. They possess a variety of spin textures topologies which could be described using  Rashba (R),  Dresselhaus  (D) and PST types. For example, we find a  combination of  R and PST (R-P) types, or  D and PST types (D-P). Among the materials investigated, MPSnBr$_3$ exhibits largest spin splitting. In addition, it already has the tendency to develop PST, suggesting that it could be a good candidate for spin texture engineering, which motivates us to focus on MPSnBr$_3$ materials in this study. MPSnBr$_3$  has been synthesized recently \cite{MPSnBr} and possesses a direct bandgap of 2.62~eV. It crystallizes in nonpolar cubic $Pm\bar3m$ phase at high temperature and undergoes a phase transition to polar orthorhombic $Pna2_1$ phase at 357~K, followed by a transition to polar monoclinic phase $Pc$  at 314~K. Above-room-temperature ferroelectricity has been predicted for this material \cite{MPSnBr, Ravi_PRM}. Computationally the ground state polarization is 3.01~$\mu$C/cm$^2$  and the  material exhibits ferroelectric Rashba effects cofunctionality\cite{Ravi_PRM}. Furthermore, it was predicted that  this material exhibit different types of spin textures  around the CBM and VBM, that is  mostly Rashba in VBM and persistent around CBM.  The \textit{\textbf{k.p}} Hamiltonian for the point group C$_s$  is 
\begin{equation} 
\label{e1}
  H =  \frac{\hbar^2}{2}\Big(\frac{k_x^2}{m_x} + \frac{k_y^2}{m_y} + \frac{k_z^2}{m_z}\Big) + \alpha k_x\sigma_y+k_y(\beta\sigma_x+\gamma \sigma_z) + \delta k_z  \sigma_y
\end{equation}
 where, $k_x$, $k_y$, and $k_z$ are components of the   k-vector in Brillouin zone and  $m_i$ (i = x, y, z) are the effective masses along Cartesian direction, $\sigma_i$ are the Pauli spin matrices. 
 The $\alpha$, $\beta$, $\gamma$ and $\delta$ are the  spin-momentum coupling parameters.  They are responsible for the type of spin textures. For example,  for the $(k_x,k_y)$ plane the Hamiltonian of Eq.~\ref{e1} yields 
 $\langle {\sigma_x} \rangle _\pm=A_\pm \cos \phi$ and $\langle {\sigma_y} \rangle _\pm=A_\pm \sin \phi$, where $A_\pm$ and $\phi$ are  the amplitude and phase, respectively, for the spin components. The $\phi$ can be estimated from $\frac{\langle {\sigma_y} \rangle_\pm}{\langle {\sigma_x} \rangle_\pm}=\frac{\alpha k_x}{\beta k_y}$. In Ref.\cite{Ravi_PRM} the coupling strengths have been reported for VB and CB. They are $\alpha$ =-0.22(0.46) eV\AA$^{-2}$, $\beta$=0.15(0.0) eV\AA$^{-2}$, $\gamma$=-0.05(-0.06) eV\AA$^{-2}$, and $\delta$=0.08(-0.11) eV\AA$^{-2}$ for the bands near VBM (CBM), respectively.  For illustrative purposes, let's focus on the VB and neglect $\gamma$ and $\delta$ parameters. Accidentally, in this case the Hamiltonian coincides with that of C$_{2v}$ point group. The spin textures occur in k$_x$-k$_y$ plane. The type of the spin textures depends on the interplay between the parameter $\alpha$ and $\beta$ and is schematically given by the phase diagram in Fig.~\ref{fig1} for k$_x$=k$_y$.  The phase diagram consists of   four parts whose boundaries are $\alpha$=$\beta$, $\alpha$=-$\beta$, $\alpha$=0, and $\beta$=0. Each region portrays different spin textures topology. For example, $\alpha$=$\beta$, predicts the Dresselhaus spin texture and $\alpha$=-$\beta$ offers Rashba type. The spin texture will be persistent for $\alpha$=0, and $\beta \neq 0$ or $\alpha \neq 0$, and $\beta$=0. There will be no spin texture for $\alpha=\beta=0$. Therefore, spin textures  engineering is achievable by tuning the coupling strengths. 

\begin{figure*}
\centering
\includegraphics[width=1.05\textwidth]{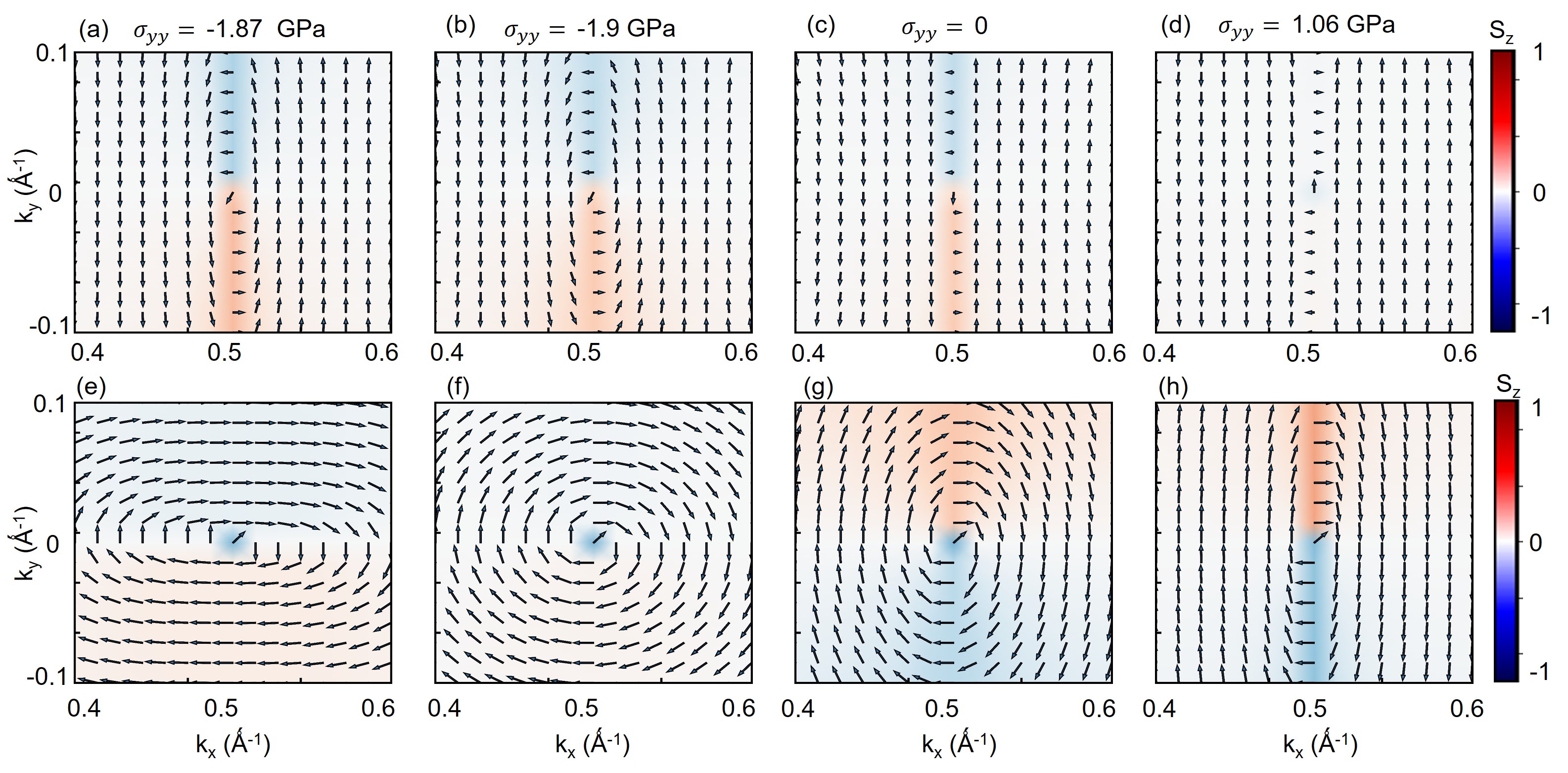}
\caption{ Near band edge spin textures of MPSnBr$_3$ in k$_x$-k$_y$ plane for different values of uniaxial stress, as given in the titles.  Here, (a)-(d) corresponds to CB and (e)-(f) corresponds to VB. Spin texture shown in (a) and (e) corresponds to stress beyond elastic limit.}
\label{fig3}
\end{figure*}

We hypothesize that it may be possible to tune these coupling through application of mechanical load.  In particular, we will test this hypothesis by application of uniaxial stress to the ground state of MPSnBr$_3$.  This models the application of uniaxial stress to the low-temperature $Pc$ phase of the material. Such stress can be induced by the force head or AFM tip  such as in piezoelectric measurements\cite{cain2014characterisation}. 
 The specific aims of this Letter are  (i)  to predict mechanical, electrical, and electronic structure response of MPSnBr$_3$ to uniaxial stress; (ii) to reveal high tunability of its spin textures by the uniaxial stress and establish its origin; (iii)  to predict stress-induced emergence of PST in this compound.

Application of uniaxial stress can be achieved in computations by constraining the lattice along one crystallographic direction, while allowing it to relax along the perpendicular ones. In this study, we  apply uniaxial stress along \textbf{b} crystallographic direction, which is perpendicular to the mirror plane, and consequently the plane of the polarization vector. Technically, this has been achieved by constraining supercell along \textbf{b} direction to the  values in the range [0.9\textbf{b}, 1.1\textbf{b}], while allowing supercell to relax along other directions. Ions were allowed to relax along all directions. This results in presence of uniaxial stress along \textbf{b} direction, which is taken to be negative of the internal stress outputted by VASP. Structural relaxation was performed using conjugate gradient algorithm with an energy convergence threshold of 0.001~meV. The residual forces on the ions were less than 14 meV/\AA\, while residual stresses along the unconstrained directions were less than 0.05GPa. Technically, we start with the ground state structure and compress (stretch) along \textbf{b}$'$ direction in steps of 0.01\textbf{b}. For each consecutive value of the \textbf{b}$'$ lattice parameter, the calculations were initialized with the structure obtained at the previous value of \textbf{b}$'$, which simulates sequential loading. The structure remained in $Pc$ space group for all values of stress investigated. To test for reversibility of the stress application, required for identifying the elastic regime, we removed the load  by retracing the steps backward. The energy and enthalpy of the  supercell as a function of stress are shown in Fig.~\ref{figS3} of Supplementary Material.

The stress-strain dependence obtained in such computations is shown in Fig.~\ref{fig2}(a). The elastic response is observed in the region of -8\% to 10\% strains corresponding to  -1.8 GPa to 1.0 GPa stresses.   To validate our computational data we estimate the Young's modulus by computing the zero stress slope of the dependence and comparing it with the data from the literature. Our computational value of  19.7 GPa,  compares well with the values of 12-30 GPa reported for other Sn and Pb based halide perovskites\cite{Young_1,young_2}.  Interestingly, under tensile load the material demonstrates rubber-like behavior associated with  early onset of nonlinear behavior \cite{Singh2021}, which is at 2 \% in this case.  Under compression the dependence remains linear down to -6 \%, although there exists a slight change of the slope at -2 \%.

\begin{figure*}
\centering
\includegraphics[width=0.9\textwidth]{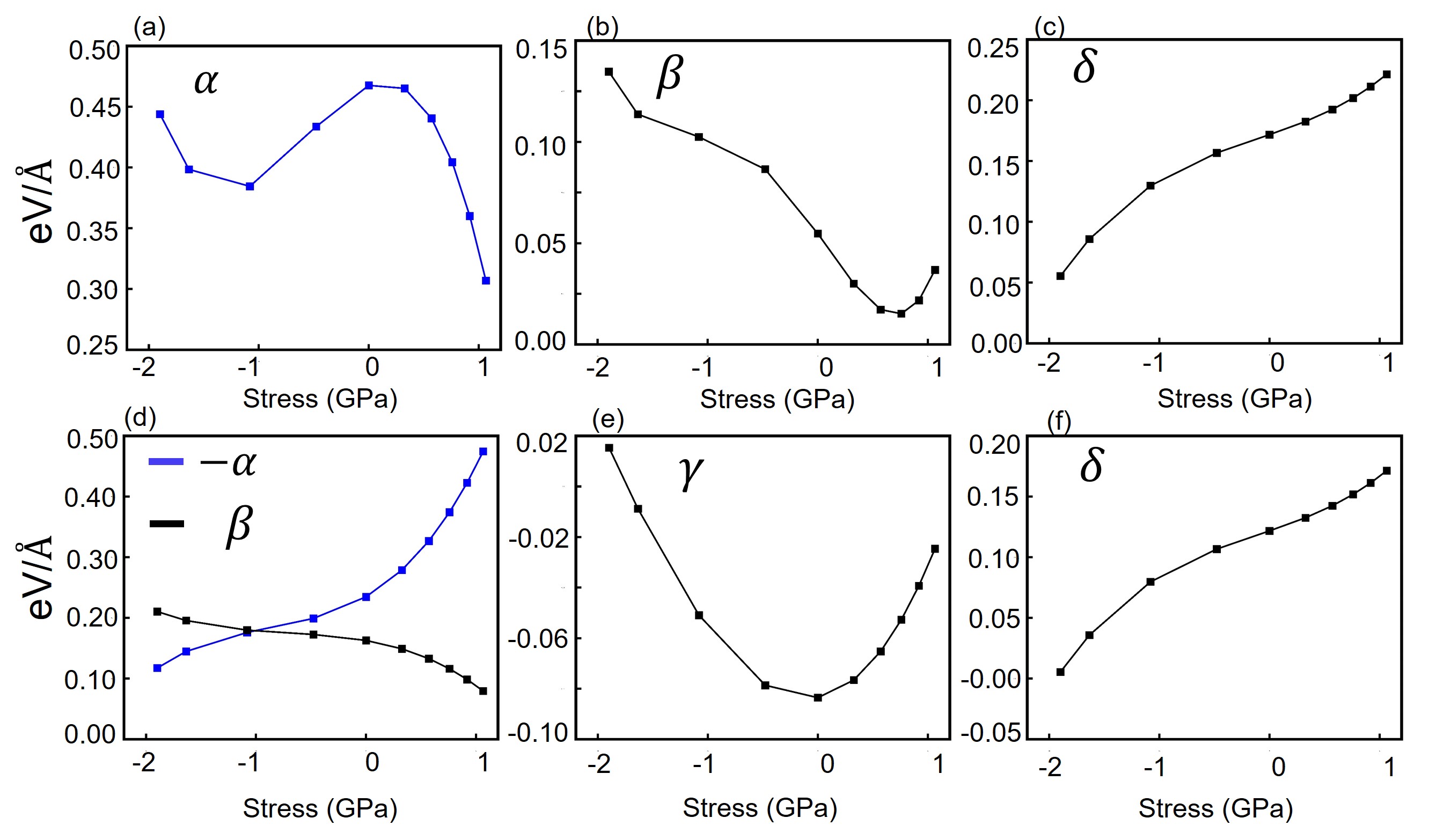}
\caption{ Dependence of spin-momentum coupling parameters on the uniaxial stress for CB [(a)-(c)] and VB[(d)-(f)]. }
\label{fig4}
\end{figure*}
To establish property tunability by application of uniaxial stress within the  elastic regime we computed polarization as a function of stress and present the data in Fig.~\ref{fig2}(b). Technically, the Berry phase approach \cite{Pola_Vanderbilt} is utilized. The data predict rather smooth  dependence of polarization on uniaxial stress. Interestingly, we find positive slope for the P$_x$ component of polarization and negative slope for P$_z$ component of polarization, which  previously has  been reported for formate perovskite\cite{Partha_AMI}. Note, that our Cartesian $x$ and $y$ axes are aligned along \textbf{a} and \textbf{b} crystallographic directions.  The associated piezoelectric strain coefficients $d_{12}$ and $d_{32}$ are 3.2 pC/N and  -2.6 pC/N, respectively. These values are comparable with TMCM-CdCl$_3$ and TMCM-MnCl$_3$  \cite{Partha_PRL}.

Next, we turn to the tunability of spin-splitting by the uniaxial stress. We have calculated the band structure along $\Gamma$-X-$\Gamma$, A-X-A and C-X-C [Here, A = (0.5, 0.5, 0), X=(0.5, 0, 0), C=(0.5, 0, 0.5)] directions of the Brillouin zone for each stress investigated. Band structures along $\Gamma$-X-$\Gamma$ direction for different stress values are shown in Fig.~\ref{figS5} of the Supplementary Material. The stress evolution of  spin splitting in both CB and VB is given in the  Fig.~\ref{fig2}(c)-(d) and predicts its high tunability by stress. For example, in the VB, tensile stress increases spin splitting along $\Gamma$-X-$\Gamma$ direction by a factor of three with respect to its stress-free value. Likewise, in the CB, compressive stress could increase spin splitting by a factor of 19 and 9 along A-X-A and C-X-C directions, respectively. Interestingly, under compressive stress the values of the spin splitting in CB along all three directions become comparable, suggesting stress-induced isotropy of this property. This should be contrasted with highly anisotropic response for the stress-free structure. 

Analysis of the electronic structure shows that in the investigated stress range  the PBE bandgap varies between 1.09eV to 1.58 eV (See Fig.~\ref{figS4} of the Supplemental Material).

Next, we turn to the investigation of the effect of uniaxial stress on spin textures.  The spin textures computed by VASP are given in Fig.~\ref{fig3}  in the k$_x$-k$_y$ planes for a few representative stresses. The spin orientation in a given k-point changes drastically under the stress. 
As already mentioned the ground state spin texture is R-P type around the VBM, and persistent around the CBM. Tensile stress converts VBM spin textures into PST with spins aligned along k$_x$ direction. Under compressive stress within elastic regime ($\sigma=$-1.9 GPa) the spin texture transforms into purely Rashba type. Going beyond elastic regime for the compressive stress ($\sigma=$-1.87 GPa) results in emergence of PST with the spins aligned along k$_x$. For all investigated stresses, the spin textures in the CB retains its persistence along the k$_y$ axis. Thus, under uniaxial stress, the ferroelectric semiconductor MPSnBr$_3$ can exhibit the PST in k$_x$-k$_y$ plane in both   CB and VB.  The spin textures in k$_x$-k$_y$ and k$_y$-k$_z$ planes for both CB and VB under different values of stress are given in Supplementary Material (See Figs.~\ref{figS6}- \ref{figS9}).

To investigate the origin of such tunability we compute the coupling strengths $\alpha$-$\delta$ for $\textbf{\textit{k.p}}$ Hamiltonian of Eq.~\ref{e1}. Technically, the eigenvalues of effective Hamiltonian are fitted with DFT band structure under stress to obtain the coupling parameters. The exact sign of the parameters is determined by comparing spin texture from $\textbf{\textit{k.p}}$ Hamiltonian with that from DFT as described in Ref.\cite{Ravi_PRM}.  The stress dependence of the coupling strengths is given in Fig.~\ref{fig4}. The data reveal high tunability of the couplings by the stress. For example, $\alpha$ changes nearly four times as we move from  compressive stress of -1.8 GPa to tensile stress of 1 GPa in the VB. Similarly, there is two times change in the $\beta$ parameter for the same stress range.  As the couplings control the type of spin textures, this explains the tunability of spin texture by the uniaxial stress found in computations. Moreover, the gradual character of the dependencies suggests a gradual evolution of the spin textures between different types, which is highly desirable for practical applications.

In summary, we utilized first-principles DFT simulations to predict that application of uniaxial stress to the low temperature polar phase of hybrid organic-inorganic perovskite MPSnBr$_3$ results in the gradual change of  the near band edge spin textures type. In particular, in case of bands in the vicinity of VBM, computations predict that the application of uniaxial tensile stress converts mostly Rashba spin textures into PST, which are highly desirable for application in spintronics. On the other hand, application of uniaxial compressive stress results in emergence of purely Rashba type spin textures followed by PST in the inelastic regime. These features originate from the tunability of spin-momentum interactions parameters by uniaxial stress.  The mechanical response of the material to uniaxial stress exhibits a rubber-like features under tensile load. Electrical response features both enhancement and suppression of  polarization components under uniaxial tension, that is both negative and positive transverse piezoelectric coefficients. The SOC induced spin splitting near band edges is also found to be  highly tunable by the uniaxial stress. For example, we found 9 to 19 fold enhancement of spin splitting  under compressive stress. The ability to manipulate properties through uniaxial stress could lead to tunable applications in spintronics.


\section*{Acknowledgements}

Authors acknowledge support from the U.S.
Department of Energy, Office of Basic Energy Sciences, Division of Materials Sciences and Engineering under Grant No. DE-SC0005245.

\bibliography{paper}

\newpage
\onecolumngrid

\begin{center}
   \textbf{\Large Supplementary Material}
\end{center}

 \vspace{1cm}

\begin{longtable*}{cccccc}
\caption{Experimentally synthesized ferroelectric hybrid organic-inorganic perovskites in C$_s$ and C$_{2v}$ crystal point groups. }\\
\hline
\endfirsthead
\multicolumn{4}{c}%
{\tablename\ \thetable\ -- \textit{Continued from previous page}} \\
\hline
Family& Compound & Space Group  &Polarization ($\mu$C/cm$^2$)& Temperature (K) & Reference\\
\hline
\endhead
\hline \multicolumn{4}{r}{\textit{Continued on next page}} \\
\endfoot
\hline
\endlastfoot
Family& Compound & Space Group  &Polarization ($\mu$C/cm$^2$)& Temperature (K) & Reference\\
\hline
Halides &MPSnBr$_3$&Cc&3.03&$<$314&\cite{MPSnBr}\\
&SAPD-PbI$_3$&Ia&10&$<$300&\cite{SAPD}\\
&BE-PbBr$_3$&Cc&3.0&$<$310&\cite{BEPbBr}\\
&TMCM-CdCl$_3$&Cc&6&383&\cite{TMCM_CdCl}\\
&TMCM-MnCl$_3$&Cc&4&406&\cite{TMCM_MnCl}\\
&[n-Budabco]CoBr$_3$&Pca2$_1$&2.4 &603 &\cite{CoBr} \\
&[(CH$_3$)$_3$NCH$_2$F]ZnCl$_3$& Pmc2$_1$&4.8 & &\cite{ZnCl}\\
& [AP]RbBr$_3$ &Ia&2.3& $<$440& \cite{RbBr}\\
& 4-ABCH-CdCl$_3$ & Pna2$_1$ &13.2&408& \cite{4-ABCH} \\

& MNCC-CdCl$_3$ & Pna2$_1$&  17.1& 363& \cite{MNCC} \\

& DMAA-CdCl$_3$ & Pna2$_1$&  1.9& 340& \cite{DMAA} \\

& 3-pl-CdCl$_3$ & Cmc2$_1$&  5.1& 316& \cite{Ye2014} \\

& 3-pl-CdBr$_3$ & Cmc2$_1$&  7& 240& \cite{Li2017} \\

\hline
Formates&DMA-Mn(HCOO)$_3$&	Cc  &2.70 - 3.61&285&\cite{DMA-Mn1}\\
&DMA-Ni(HCOO)$_3$&	Cc &0.42 - 0.52&285&\cite{DMA-Ni}\\
&DMA-Co(HCOO)$_3$&	Cc &0.30&285&\cite{DMA-Co}\\
&DMA-Zn(HCOO)$_3$&	Cc &0.45&285&\cite{DMA-Zn}\\
&HAZ-Mn(HCOO)$_3$&	Pna2$_1$ &3.58&110&\cite{HAZ-M}\\
&HAZ-Co(HCOO)$_3$&	Pna2$_1$ &2.61&405&\cite{HAZ-M}\\
&HAZ-Zn(HCOO)$_3$&	Pna2$_1$ &3.48&110&\cite{HAZ-M}\\
&HAZ-Mg(HCOO)$_3$&	Pna2$_1$ &3.44&400&\cite{HAZ-M}\\
&EA-Mg(HCOO)$_3$&	Pna2$_1$ &3.43&93&\cite{EA-Mg}\\
&Gua-Cu(HCOO)$_3$&	Pna2$_1$ &0.22 - 0.37&293&\cite{Gua-Cu1, Gua-Cu2}\\
\hline
2D halide &BA$_2$PbCl$_4$ &	Cmc2$_1$&2.1&328&\cite{BAPbCl}\\
  &MHy$_2$PbBr$_4$ &	Pmn2$_1$&4.6–5.8 &350&\cite{MePbBr}\\
 &AMP$_2$PbI$_4$ &	Pc& 9.8& 353&\cite{AMPPbI}\\
 &BZA$_2$PbCl$_4$ &	Cmc2$_1$&13 & 438 &\cite{BZA}\\
 &ATHP$_2$PbBr$_4$ &	Cmc2$_1$&5.6 &503&\cite{ATHP}\\
 &CHA$_2$PbBr$_4$	 &Cmc2$_1$&6.6 & 364 &\cite{CHA}\\
 &FBZA$_2$PbCl$_4$ &	Cmc2$_1$&5.35 &448&\cite{FBZA}\\
 &pFBZA$_2$PbBr$_4$ &	Cmc2$_1$&4.2 & 440&\cite{pFBZA}\\
 &DFCHA$_2$PbI$_4$ &	Cmc2$_1$& 4.5 & 377&\cite{DFCHA}\\

\hline

Cyanide & (TrMNO)$_2$[KFe(CN)$_6$] & Cc &    1.25 &$<$408 & \cite{Cyanide}\\

\end{longtable*}
MP = Methylphosphonium (CH$_3$PH$_3$),
SAPD = 1-((2-hydroxybenzylidene)amino)pyridin-1-ium,\\ BE=BrCH$_2$CH$_2$N(CH$_3$)$_3$
DMAA = N,N-dimethylallylammonium,  4-ABCH =  4-fluoro-1-azabicyclo[2.2.1]heptane, MNCC = Me$_3$NCH$_2$CH$_2$OH, DMA = (CH$_3$)$_2$NH$_2$,
3-pl=3-Pyrrolinium, HAZ=NH$_2$NH$_3$,
Gua=C(NH$_2$)$_3$, BA=(C$_4$H$_9$NH$_3$)$_2$, BZA=benzylammonium, MHy = CH$_3$NH$_2$NH$_{2}$, AMP=4-(aminomethyl)piperidinium, ATHP=4-aminotetrahydropyran, CHA=cyclohexylaminium, FBZA=ﬂuorobenzylammonium, DFCHA = 4,4-difluorocyclohexylammonium
TrMNO = (CH$_3$)$_3$NOH
 
 \renewcommand{\thefigure}{S\arabic{figure}}

\setcounter{figure}{0}

 \begin{figure*}
\centering
\includegraphics[width=0.9\textwidth]{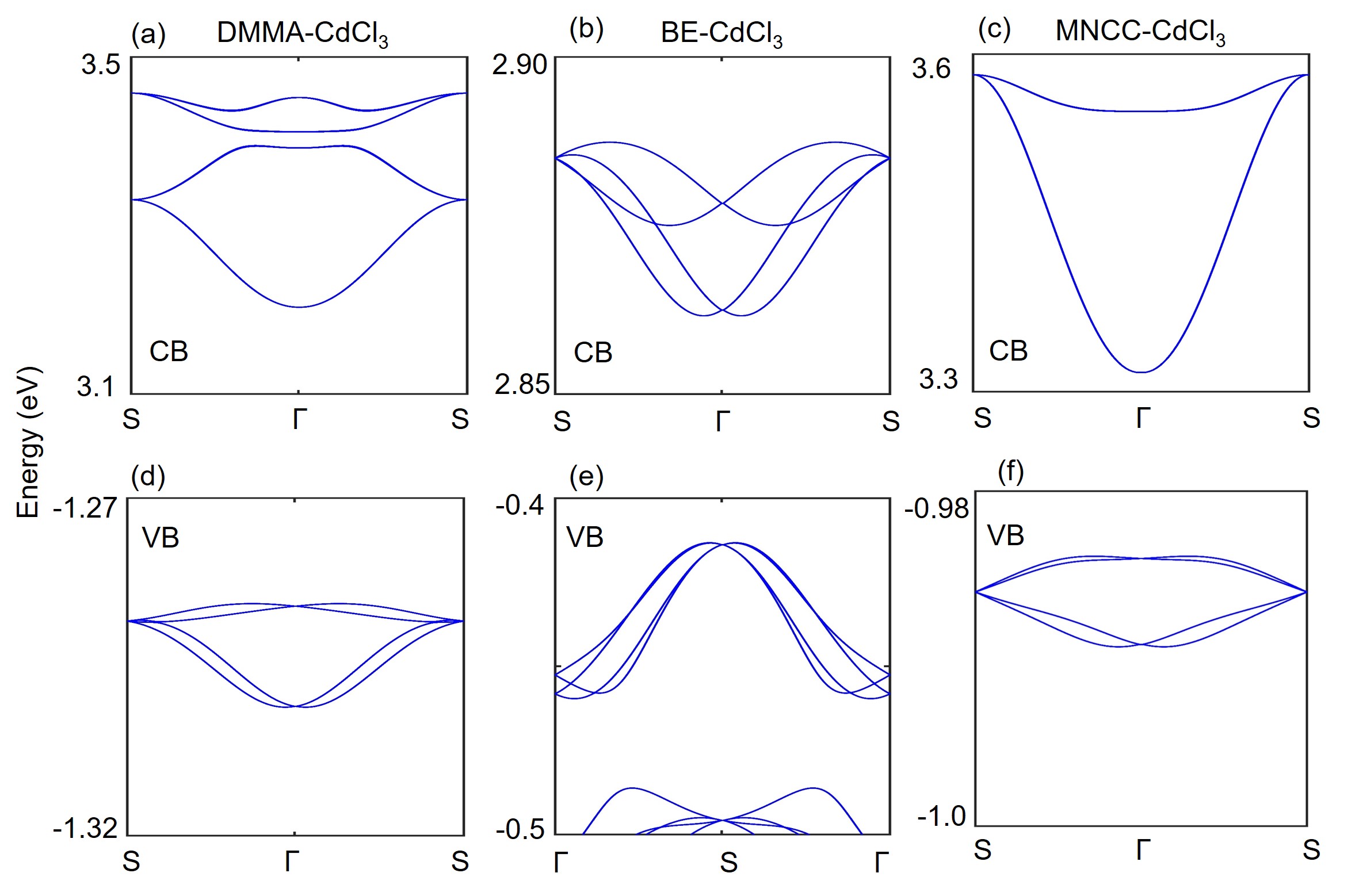}
\caption{ Band structure of some hybrid organic and inorganic perovskites listed in Table 1.     }
\label{figS1}
\end{figure*}

\begin{figure*}
\centering
\includegraphics[width=0.9\textwidth]{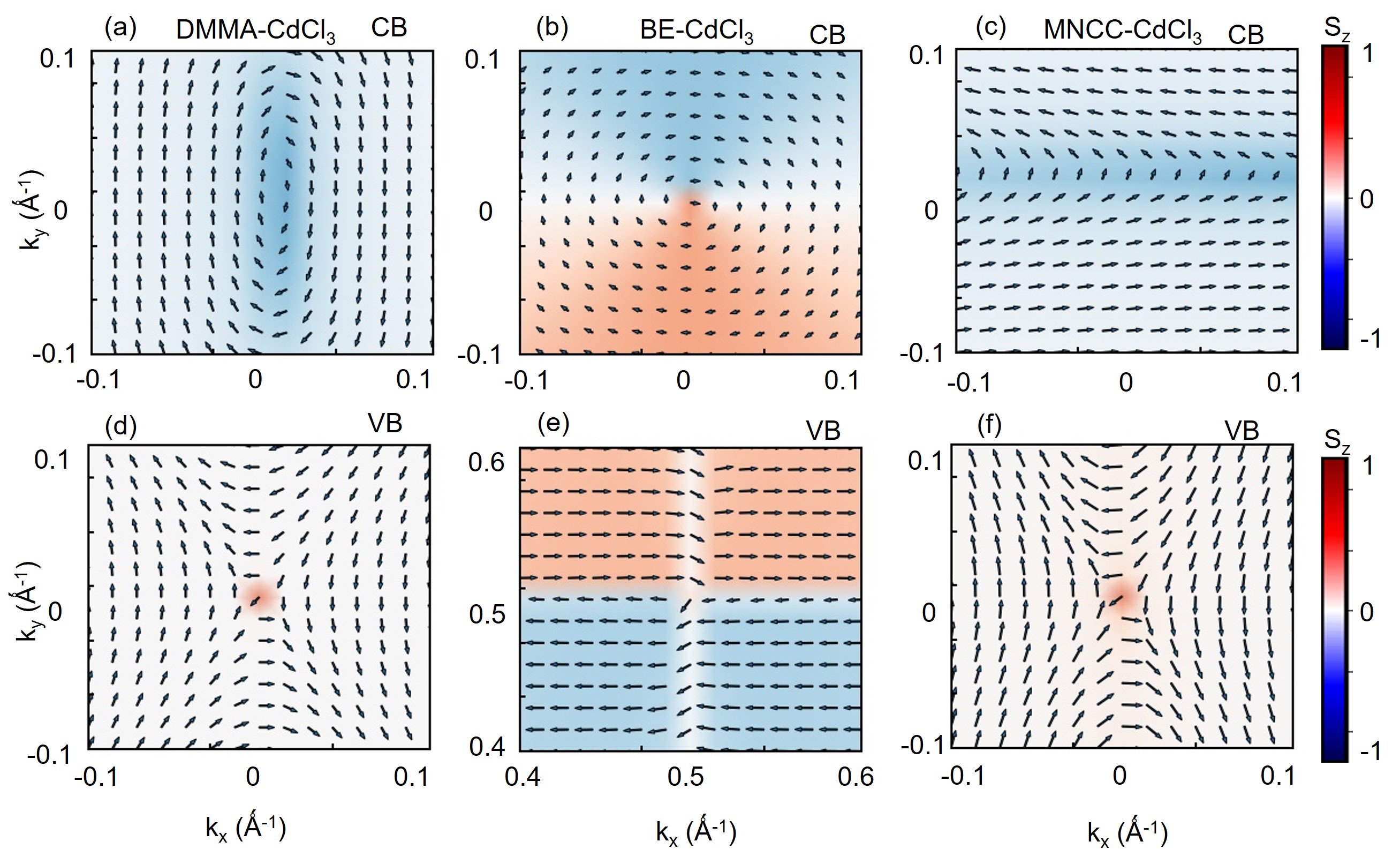}
\caption{ Spin structure of some hybrid organic and inorganic perovskites listed in Table 1.    }
\label{figS2}
\end{figure*}

\begin{figure*}
\centering
\includegraphics[width=0.7\textwidth]{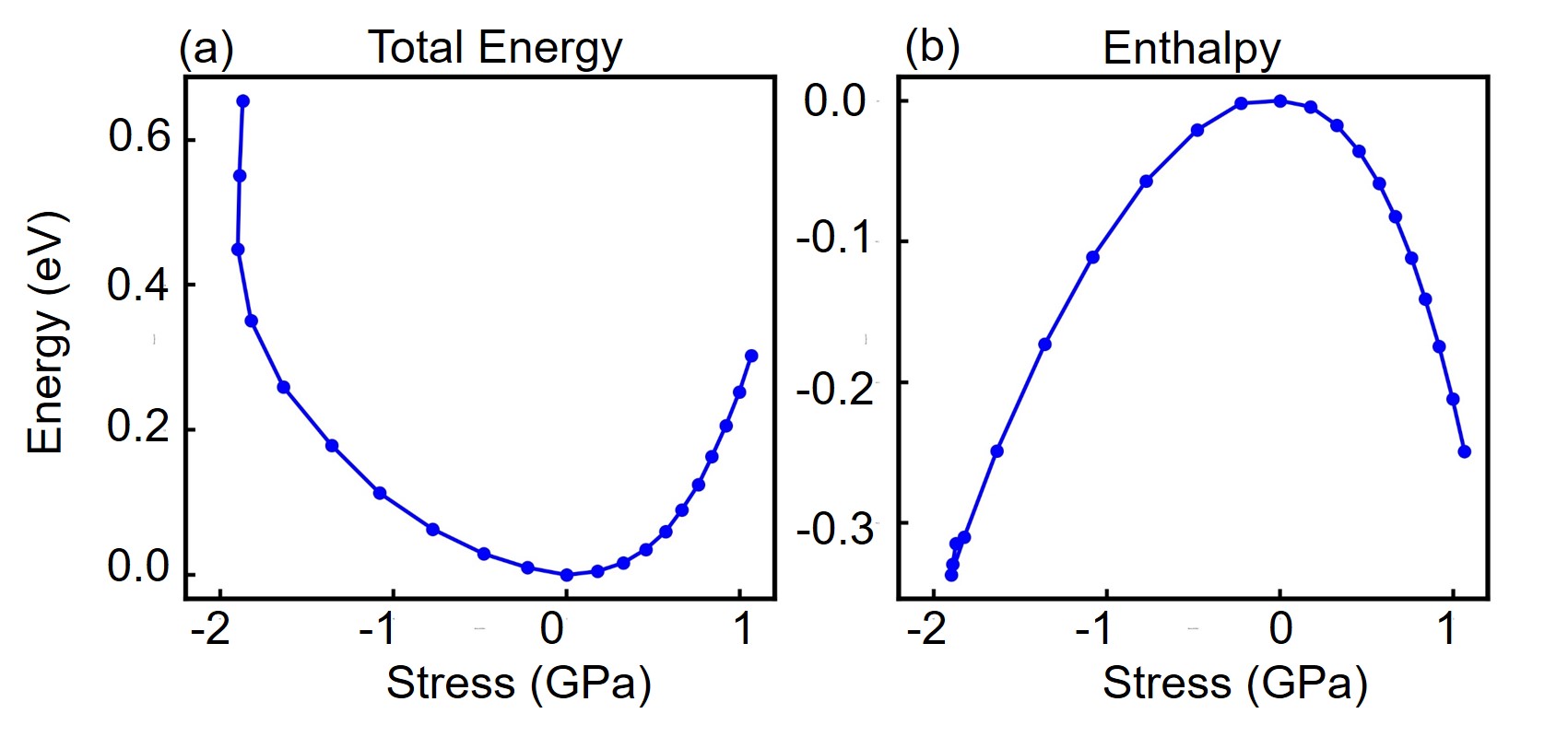}
\caption{  Variation of total energy and enthalpy as a function of uniaxial stress.     }
\label{figS3}
\end{figure*}

\begin{figure*}
\centering
\includegraphics[width=0.7\textwidth]{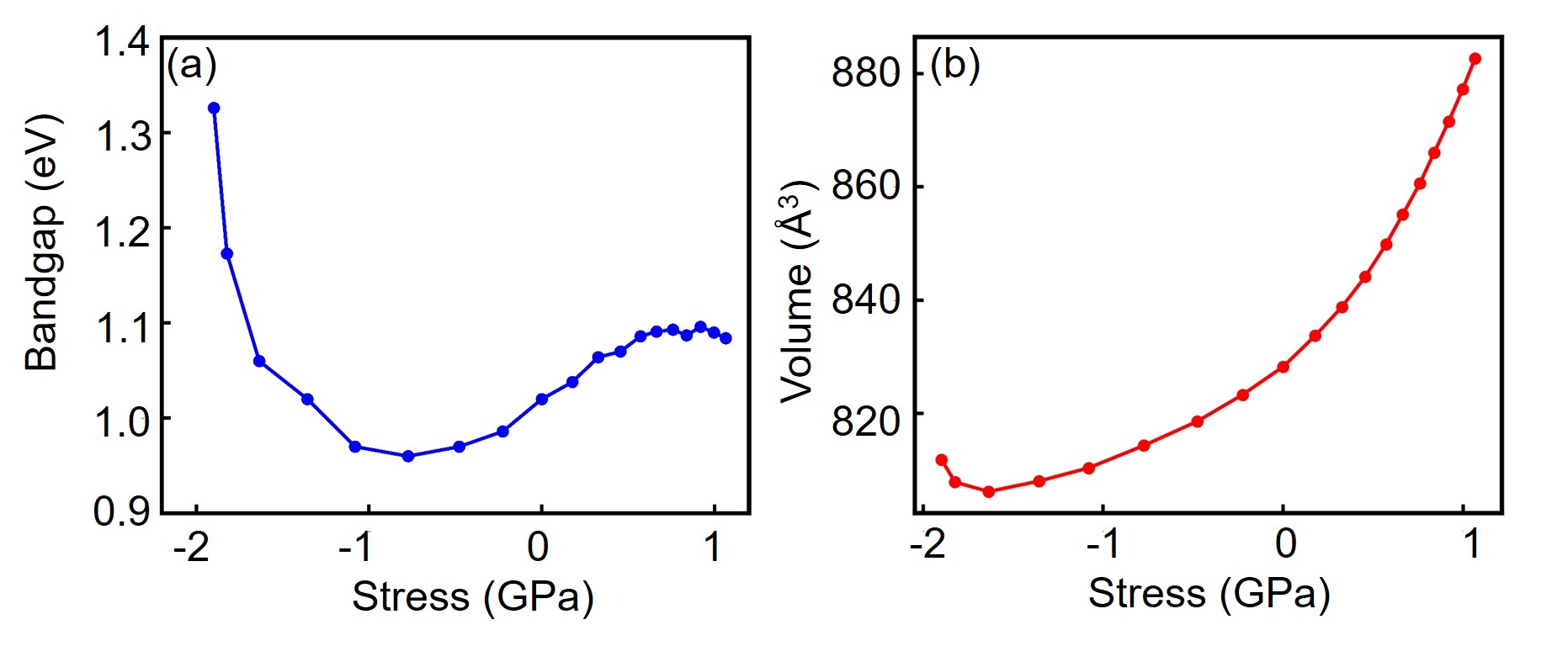}
\caption{  Variation of electronic bandgap and volume as a function of uniaxial stress.     }
\label{figS4}
\end{figure*}

\begin{figure*}
\centering
\includegraphics[width=1.\textwidth]{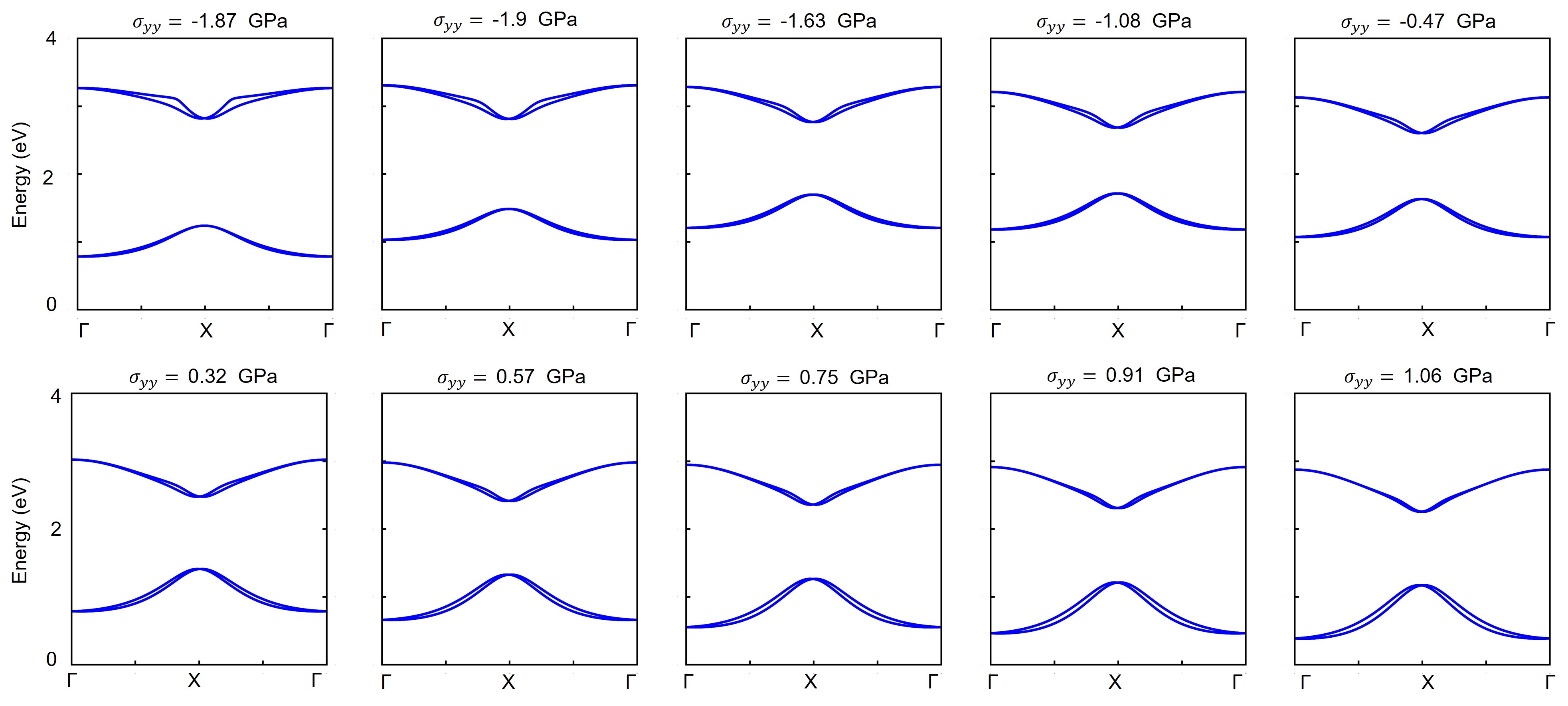}
\caption{ Band structure MPSnBr$_3$ under uniaxial stress.     }
\label{figS5}
\end{figure*}

\begin{figure*}
\centering
\includegraphics[width=1.1\textwidth]{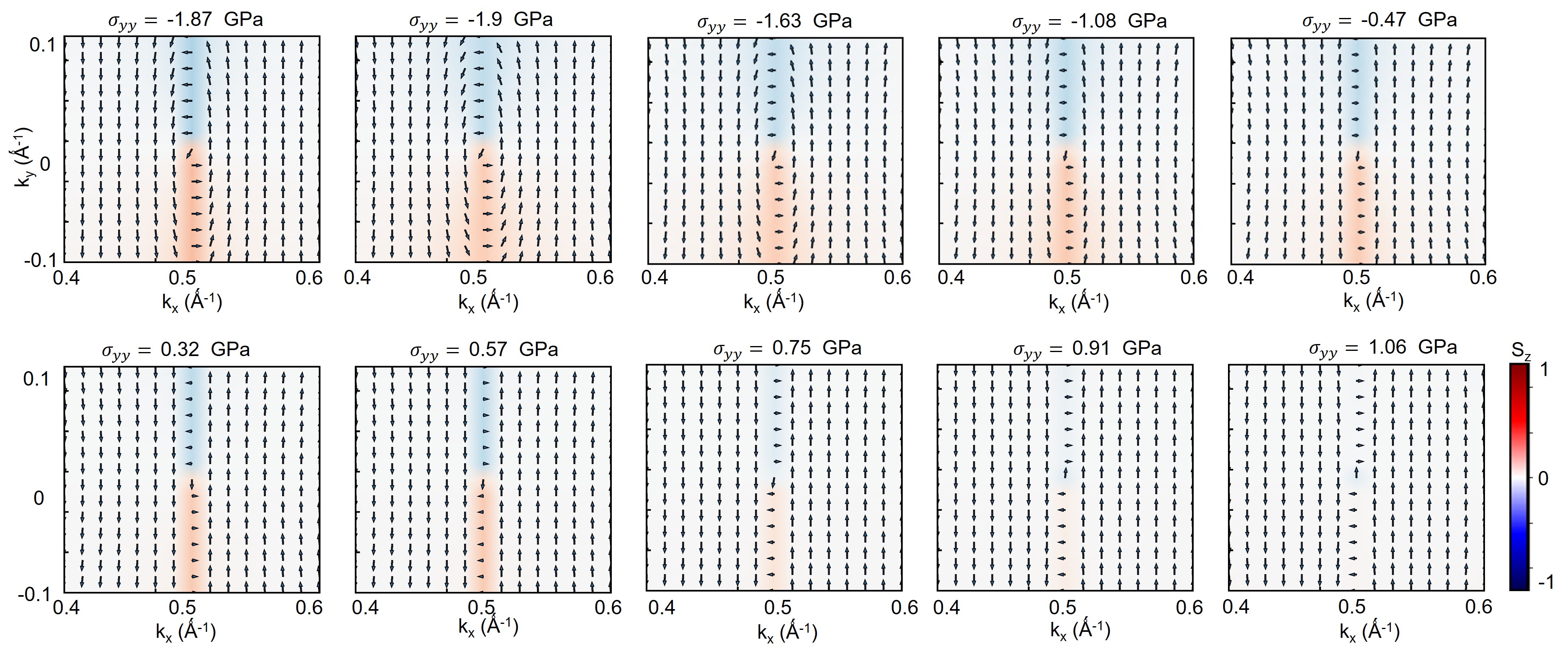}
\caption{  Spin textures of MPSnBr$_3$ lower most conduction band in k$_x$-k$_y$ plane under uniaxial stress.    }
\label{figS6}
\end{figure*}

\begin{figure*}
\centering
\includegraphics[width=1.1\textwidth]{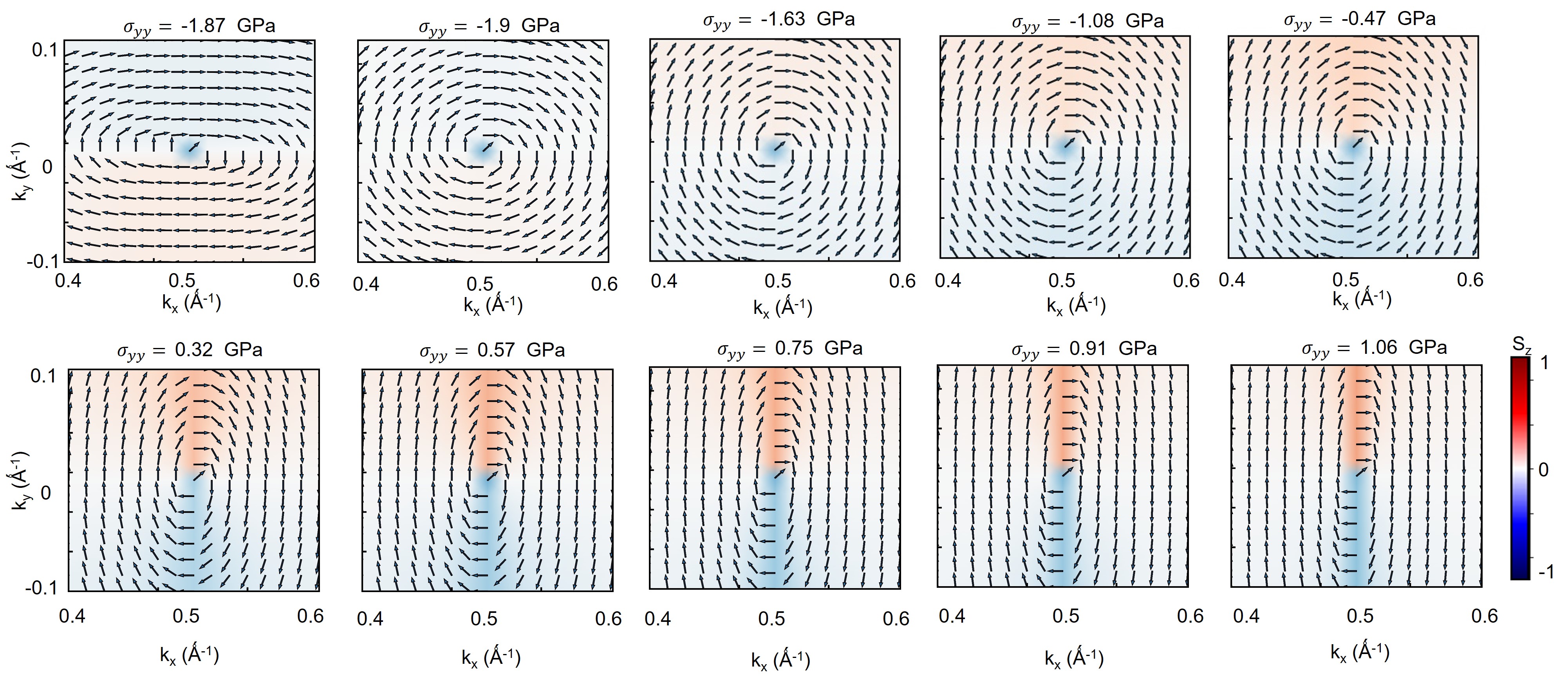}
\caption{  Spin textures of MPSnBr$_3$ upper most valence band in k$_x$-k$_y$ plane under uniaxial stress.    }
\label{figS7}
\end{figure*}

\begin{figure*}
\centering
\includegraphics[width=1.1\textwidth]{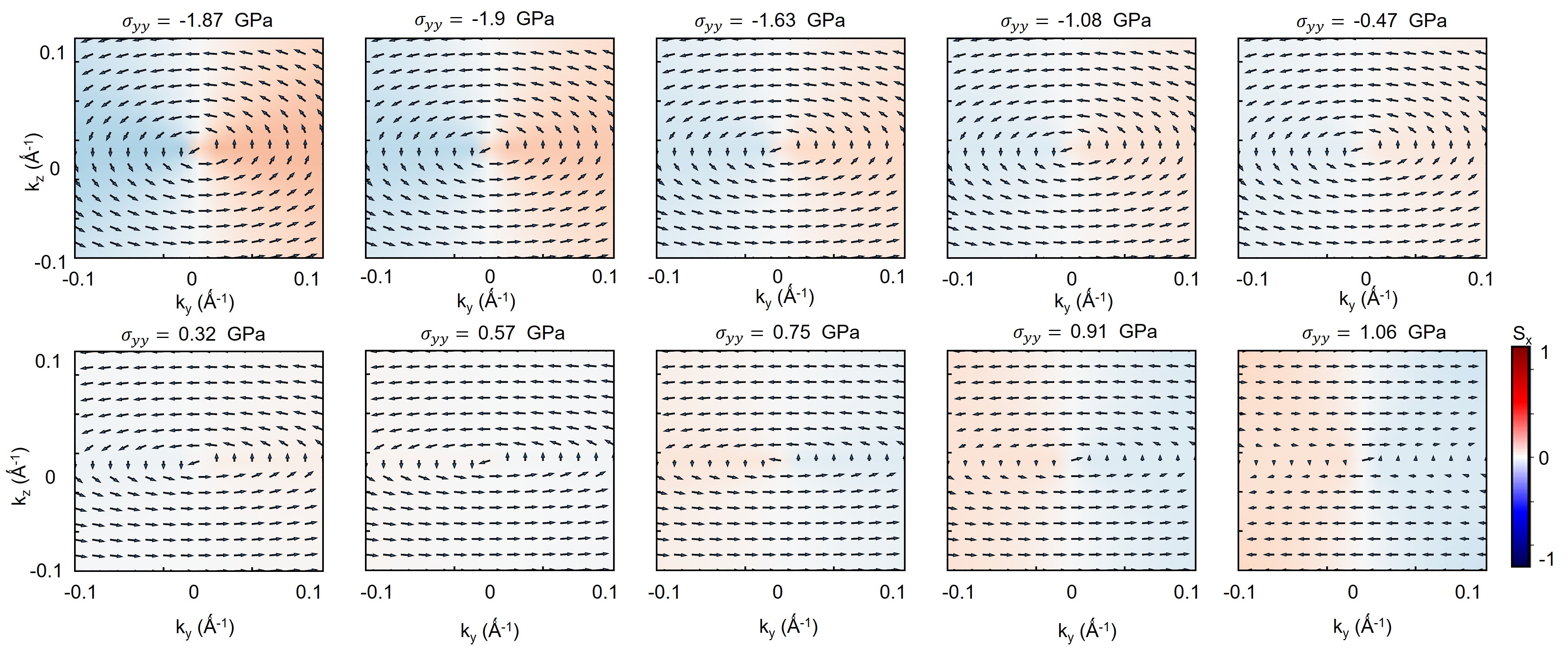}
\caption{  Spin textures of MPSnBr$_3$ lower most conduction band in k$_y$-k$_z$ plane under uniaxial stress.    }
\label{figS8}
\end{figure*}

\begin{figure*}
\centering
\includegraphics[width=1.1\textwidth]{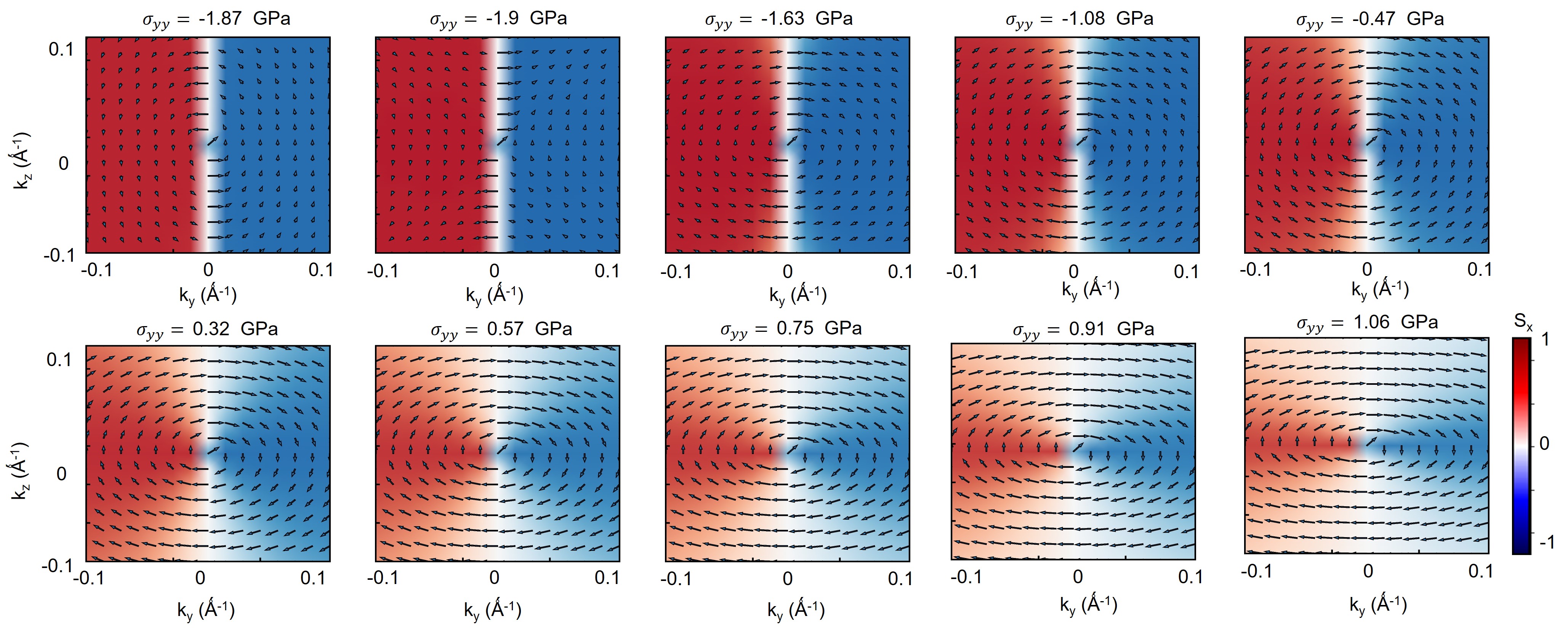}
\caption{  Spin textures of MPSnBr$_3$ upper most valence band in k$_x$-k$_y$ plane under uniaxial stress.    }
\label{figS9}
\end{figure*}

\end{document}